# The Impact of Student Writing Assessment Literacy on Psychological Factors: An Ordinal Logistic Regression Analysis

Siyi Cao[12*/**], Linping Zhong[1*], Chao Wang[13**]

1. School of Foreign Languages, Southeast University, Nanjing, China
2. Department of Chinese and Bilingual Studies, The Hong Kong Polytechnic University, Hong Kong, China
3. Graduate School of Arts and Sciences, the University of Tokyo

* Equal contribution, sharing first authorship
** Corresponding authors: siyi.c@nuaa.edu.cn, phoenix--wong@hotmail.com

**Abstract**

Previous studies have shown that enhanced student assessment literacy can lead to improvements in academic performance in EFL (English as a Foreign Language) writing. Additionally, psychological factors such as self-efficacy, achievement motivation, and writing anxiety significantly influence EFL writing outcomes. However, the relationship between student writing assessment literacy (SWAL) and these psychological factors remains unclear. The present study aims to explore how SWAL affects these psychological factors in the Chinese EFL context. Data were collected from 103 Chinese undergraduate EFL students using four questionnaires: the Student Writing Assessment Literacy Scale (SWAL), the Self-Efficacy for Writing Scale (SEWS), the Achievement Goal Questionnaire (AGQ), and the Second Language Writing Anxiety Inventory (SLWAI). Ordinal logistic regression was employed to analyze the data. The results indicated that higher levels of SWAL were positively associated with writing self-efficacy and achievement motivation, while negatively related to writing anxiety. These findings have significant pedagogical implications for second language (L2) writing instructors, emphasizing the importance of integrating SWAL training into writing instruction to enhance students' writing experiences and outcomes.

**Keywords**: student writing assessment literacy, self-efficacy, achievement motivation, and writing anxiety, L2 writing, EFL learners

## 1. Introduction

Assessment literacy, the knowledge and skills needed to effectively achieve educational goals and meet assessment objectives, has aroused great attention in recent years (Fulcher, 2012; Popham, 2009; Taylor, 2009). Evidence from literature demonstrated that students play an important role in the process of assessment and their assessment literacy can transform practices in higher education by encouraging greater responsibility for their own learning (Xu & Zheng, 2024). In addition, research suggests that increased student assessment literacy can lead to improvements in academic performance (e.g., Hannigan et al., 2022). In order to test student assessment literacy in general education settings, Chan and Luo (2021) developed and validated a scale in terms of four dimensions: knowledge, attitude, action, and critique.

Student assessment literacy within the context of second language (L2) writing is also important for students because students grasp the criteria for writing assessment, take ownership of and evaluate their work, and make informed decisions about their writing efforts (Huot, 2002). Although Xu et al. (2023) created a scale of student writing assessment literacy (SWAL) in the L2 writing context, it is still unclear whether SWAL affects internal psychological factors, such as self-efficacy, achievement motivation and writing anxiety.

Psychological factors, such as self-efficacy, achievement motivation, and writing anxiety, significantly impact EFL academic performance (Sabti et al., 2019). Research indicates that high achievement motivation and writing self-efficacy can substantially enhance writing performance, while writing anxiety tends to hinder writing ability (Pajares et al., 2006; Schunk & Swartz, 1993; Zhang & Guo, 2012). Although increased student assessment literacy has been shown to improve academic performance, the relationship between SWAL and these psychological factors remains unclear. To address this gap, the present study examines the connections between SWAL and self-efficacy, achievement motivation, and writing anxiety among Chinese EFL undergraduate learners.

## *1.1 Student assessment literacy in the L2 writing context*

Student assessment literacy is reported to be effective to academic learning because students can clearly understand the goals and methods behind assessments, along with the strengths and weaknesses of their own work, which they can gain greater control over their learning process (Lodge, 2008; Popham, 2009; Sadler, 2009). Initially, Smith et al. (2013) proposed a model for conceptualizing student assessment literacy, identifying three key components: understanding the purpose of assessment, evaluating one's own responses, and recognizing the assessment processes. Building on this framework, some studies like Torshizi and Bahraman (2019) found that students who took on quasi-teacher roles, i.e., providing English lessons to their peers, were more assessment literate than their counterparts. Additionally, student assessment literacy can be influenced by various factors, such as targeted training or interventions. According to Smith et al. (2013), professional development programs could significantly enhance student assessment literacy, though the gap between theory and practice still exists in some educational settings.

In the realm of second language (L2) writing, Xu et al. (2023) developed and validated a comprehensive scale to assess student writing assessment literacy (SWAL), incorporating four dimensions: knowledge, belief, behavior, and critique. The knowledge dimension involves understanding assessment methods, their purposes, and potential drawbacks. The belief dimension relates to students' perceptions of how writing assessments affect their skills, emotions, and engagement. The behavior dimension covers strategies for writing tasks, self-reflection, and using feedback for improvement. The critique dimension focuses on critical thinking and discussions about assessment results and feedback. Based on this new scale, a few empirical studies explored SWAL from various perspectives. For example, Özçubuk and Merç (2024) conducted a study at a state university in Turkey, using semi-structured interviews, and found a significant correlation between students' understanding of assessment objectives and their positive attitudes toward learning. Xu and Zheng (2024) studied changes in SWAL and L2 writing engagement over a semester. While SWAL improved, there was no corresponding increase in writing engagement, though the behavior dimension of literacy was found to predict engagement. Despite these advancements, the impact of SWAL on important psychological factors related to L2 writing remains unclear.

## 1.2 *Self-efficacy in the L2 writing context*

Psychological factors, such as self-efficacy, are also considered crucial in influencing writing performance. Self-efficacy refers to an individual's belief in their own abilities, which plays a pivotal role in motivating behavior (Bandura &Wessels, 1997). To illustrate, self-efficacy serves as a significant predictor of how an individual acts in the writing task. It is reported that a high level of writing self-efficacy reflects a strong sense of confidence when completing writing tasks. For example, Schunk and DiBenedetto (2016) claimed that individuals with elevated self-efficacy in writing are likely to show greater enthusiasm and exert more effort when engaging in writing activities (Schunk & Usher, 2012). Moreover, high self-efficacy fosters perseverance and resilience when encountering challenges during writing tasks. For instance, Pajares and Johnson (1996) found that among various motivational factors, self-efficacy frequently emerged as the strongest predictor.

In the field of second language (L2) writing, numerous studies across different languages have demonstrated the influence of self-efficacy on writing performance among EFL learners. For Iranian EFL learners, Rahimpour and Nariman-Jahan (2010) examined the impact of self-efficacy on writing performance across dimensions such as conceptual load, fluency, complexity, and accuracy. Their findings indicated a significant correlation between self-efficacy and conceptual load. Similarly, Zhang and Guo (2012) conducted a study involving 66 Chinese EFL learners and found a positive and significant relationship between self-efficacy and English writing proficiency among freshmen majoring in English, though this correlation did not extend to sophomores. For Korean EFL learners, Park (2024) reported a positive association between writing self-efficacy and proficiency levels, with learners grouped into high and low proficiency categories based on their writing scores. However, it is still unclear about the relationship between self-efficacy and SWAL related to L2 writing to date.

## 1.3 *Achievement motivation in the L2 writing context*

Achievement motivation refers to an individual's intrinsic drive to excel, fueled by a sense of personal accomplishment (Elias et al., 2010; Singh, 2011). Research has consistently shown that achievement motivation plays a vital role in enhancing academic performance, with a strong positive correlation between motivation and academic success. It is widely regarded as a dynamic and powerful predictor of academic outcomes (Chea & Shumow, 2017; Emmanuel et al., 2014; Tamannaifar & Gandomi, 2011). In particular, high achievement motivation, when coupled with strong self-efficacy, often leads individuals to invest considerable effort in achieving their goals in tasks to which they are committed (Elias et al., 2010).

Achievement motivation also plays a critical role in the context of second language (L2) writing. Numerous studies have demonstrated a significant positive correlation between achievement motivation and writing performance. For instance, Sabit et al. (2019) found that Iraqi EFL learners with higher levels of writing achievement motivation tend to perform better in writing tasks. This finding is supported by Nasihah and Cahyono (2017) in their study of Indonesian EFL students. However, Wang (2021) reported that achievement motivation does not predict creative writing performance. Despite these insights, no research has yet explored whether strategies such as Self-Writing and Learning (SWAL) can influence achievement motivation in the L2 writing context.

### 1.4 Writing Anxiety in the L2 writing context

Anxiety refers to the the experience of unease, worry, and physiological responses that a learner encounters when engaging in tasks (Cheng, 2004). In the context of writing, anxiety is considered as a situational phenomenon, where feelings of concern are coupled with physical reactions such as excessive sweating, rapid heartbeat, and negative expectations (Sabit et al., 2019). It was found that anxiety is a significant factor influencing academic performance (e.g., Mirawdali et al., 2018). Specifically, anxiety has been shown to negatively impact the writing outcomes of L1 learners (e.g., Daly & Miller, 1975). Furthermore, it was reported that anxiety is associated with other factors, including self-efficacy and motivation (Sabit et al., 2019).

Regarding EFL learners, writing anxiety is generally regarded as detrimental to students' writing performance. For example, a study by Zhang (2011) revealed a negative correlation between writing anxiety and writing performance among Chinese English majors. This may be attributed to factors such as linguistic difficulties, insufficient writing practice, test anxiety, lack of topical knowledge, and low self-confidence. Similarly, research by Negari and Rezaabadi (2012) found a significant correlation between anxiety and final writing test performance among Iranian EFL learners. However, they also suggested that by leveraging the facilitative aspects of anxiety, students could potentially improve their writing performance. Furthermore, it has been reported that peer feedback can help reduce writing anxiety during the writing process (Bolourchi & Soleimani, 2021).

In summary, previous studies have highlighted three key points relevant to the current topic. First, increased student assessment literacy has been shown to enhance academic performance. Second, writing self-efficacy and writing achievement motivation are significant factors that can greatly improve writing performance. Third, writing anxiety has been found to negatively impact writing ability. Despite these insights, it remains unclear whether student writing assessment literacy (SWAL) influences psychological factors such as writing self-efficacy, writing achievement motivation, and writing anxiety. Therefore, the present study seeks to address the following research questions:

RQ1: Does SWAL influence writing self-efficacy ?
RQ2: Does SWAL influence achievement motivation?
RQ3: Does SWAL influence writing anxiety?

## 2. Method

### 2.1 Participants

The sample for this study included 103 Chinese EFL freshmen (57 female, 46 male) from two classes at a first-tier university in China. All participants voluntarily took part in the study. Their ages ranged from 18 to 24 years ($M = 21.36$, $SD = 1.23$). Each participant had scored above 90 on the National College Entrance Examination (NCEE), which has a maximum score of 150, indicating that they possessed a middle level of English proficiency. The participants' first language was Chinese, and they were learning English as a foreign language. According to self-reports, none of them had any disorders with speaking or hearing.

The research design followed the paradigm outlined by Sabti et al. (2019), and the study was approved by the Human Research Ethics Committee of the university affiliated with the first author.

### 2.2 Questionnaires

In order to address these three research questions, this study used four questionnaires. Firstly, student writing assessment literacy (SWAL) scale, created by Xu et al. (2023), served as a tool to measure the understanding of students in terms of the purpose and process involved in the writing assessment. This SWAL scale comprised four sub-components with totally 25 items (5-point Likert), i.e., knowledge, belief, behavior and critique.

Secondly, to examine students' self-efficacy, this study utilized the self-Efficacy for writing scale (SEWS) designed by Bruning et al. (2013). The scale comprises 16 items divided into three sub-scales (5-point Likert), covering the writing ideation, the writing conventions and the writing self-regulation.

Thirdly, the achievement goal questionnaire (AGQ), created by Elliot and Church (1997), was adopted to gather data on students' achievement motivation. This questionnaire includes three sub-dimensions of 18 items (7-point Likert): mastery-approach goals, performance-approach goals, and performance-avoidance goals.

Fourthly, to access students' writing anxiety, the Second Language Writing Anxiety Inventory (SLWAI) developed by Cheng (2004) was utilized. Three sub-components with 22 items (5-point Likert) were contained, i.e., cognitive anxiety, somatic anxiety, and avoidance behavior anxiety.

This study involved translating the four questionnaires aforementioned into Chinese, aiming to improve participants' understanding of the items and provide a helpful tool for future researchers in the field. The final translation was checked by two professors in linguistics and carefully scrutinized to ensure it accurately reflected the original instruments.

*2.3 Procedure*

The study was conducted in two classroom settings, with 52 participants in one and 51 in the other. To collect the data, the first author collaborated with an English teacher who was tutoring two classrooms at a first-tier university In China. To control for the potential influence of the semester period on SWAL, which was found to remain at a moderate to high level from the beginning to the end of the semester (Xu & Zheng, 2024), data were collected at the end of the semester, just before the final English writing assessment (after the teacher had distributed the assessment rubric). In practice, the teacher instructed the students to complete all the questionnaires. None of the participants were informed of the study's purpose. Additionally, the teacher assured them that their responses would remain confidential and discarded after statistical analysis, and would not affect their final test scores. Participants were also made aware that they could choose to decline or withdraw from the survey at any time. To protect privacy, no personal identifiers, such as names, were collected. Each survey took approximately 6 minutes to complete.

*2.4 Data Analysis*

There are two stages for the data analysis conducted in this study. In the first stage, descriptive statistics were used to categorize the average scores of three key factors (i.e., writing self-efficacy, writing achievement motivation, and writing anxiety) into three levels (e.g., low, moderate, high) based on participants' responses to the corresponding questionnaires. Regarding the categorization criteria, we deviated from the approach used by Sabti et al. (2019), which relied solely on fixed cutoff points (e.g., 1-2.99 as low level) without considering the distribution of the data or potential variability across different samples. In contrast, we defined the categories based on percentile thresholds (e.g., Nazemosadat & Ghasemi, 2004): responses below the 30th percentile were categorized as low, those between the 30th and 70th percentiles as moderate, and those exceeding the 70th percentile as high.

The second stage employed ordinal logistic regression, as the independent variable was continuous (the average SWAL score), while the dependent variable was categorical, representing the different levels of the three factors (writing self-efficacy, writing achievement motivation and writing anxiety). To assess whether SWAL significantly affected the different levels of each factor, the study used the "statsmodels" package in Python 3.5 to perform the ordinal logistic regression.

## 3. Results

### 3.1 SWAL vs. writing self-efficacy

Table 1 presented the levels of writing self-efficacy across different groups. The mean scores for writing self-efficacy indicate distinct differences between the groups. Specifically, the mean writing self-efficacy score for the low group was 3.19 ($SD = 0.37$), for the moderate group was 3.97 ($SD = 0.19$), and for the high group was 4.50 ($SD = 0.17$). These results suggest a clear trend, with individuals in the high self-efficacy group reporting significantly higher levels of writing confidence compared to those in the moderate and low self-efficacy groups.

In order to assess the relationship between SWAL and writing self-efficacy, an ordinal logistic regression was performed. The full model was statistically significant ($\chi^2(100) = 87.22, p < .001$). The results revealed that SWAL was a strong predictor of writing self-efficacy ($B = 5.97, SE = .91, p < .001$). Specifically, a one-unit increase in SWAL significantly increased the odds of being classified into a higher self-efficacy category.

Additionally, the categorical predictor of self-efficacy (low/moderate vs. moderate/high) had a significant effect ($B = 21.89, SE = 3.46, p < .001$). Participants in the low/moderate self-efficacy category had exceptionally higher odds of being in a higher self-efficacy category compared to those in the moderate/high category. Furthermore, participants in the moderate/high self-efficacy group were more likely to be in a higher self-efficacy category than those in the low/moderate group ($B = 1.20, SE = 0.15, p < .001$).

The probabilities of being in different self-efficacy categories across levels of SWAL are visualized in Figure 1a. Participants with lower SWAL scores are predominantly in the low self-efficacy category. As SWAL scores increase, the likelihood of being in the moderate self-efficacy category rises, peaking around mid-level SWAL scores. At higher SWAL scores, the high self-efficacy category becomes dominant. These trends visually reinforce the regression findings, highlighting the strong positive relationship between SWAL and self-efficacy levels.

Table 1. Levels of Writing self-efficacy

| Groups of writing self-efficacy | Mean | SD |
|---|---|---|
| Low | 3.19 | .37 |
| Moderate | 3.97 | .19 |
| High | 4.50 | .17 |

Table 2. Ordinal Logistical Regression Results for SWAL predicting Writing Self-efficacy, Motivation, and Anxiety

| | B | SE | z | p | 95% CI |
|---|---|---|---|---|---|
| *SWAL × self-efficacy* | | | | | |
| SWAL | 5.97 | .91 | 6.58 | .000 | [4.19, 7.74] |
| low/moderate | 21.89 | 3.46 | 6.33 | .000 | [15.10, 28.67] |
| moderate/high | 1.20 | .15 | 7.97 | .000 | [ .91, 1.50] |
| *SWAL × motivation* | | | | | |
| SWAL | 1.82 | .46 | 3.99 | .000 | [ .93, 2.72] |
| low/moderate | 6.14 | 1.77 | 3.46 | .001 | [2.66, 9.62] |
| moderate/high | .66 | .14 | 4.75 | .000 | [ .39, .94] |
| *SWAL × anxiety* | | | | | |
| SWAL | -3.32 | .59 | -5.61 | .000 | [-4.48, -2.16] |
| low/moderate | -14.34 | 2.44 | -5.88 | .000 | [-19.13, -9.56] |
| moderate/high | .86 | .14 | 6.04 | .000 | [ .58, 1.14] |

### *3.2 SWAL vs. writing achievement motivation*

Table 3 displayed the writing achievement motivation levels across the different groups. The results reveal notable variations in the mean motivation scores between the groups. For the low motivation group, the mean score was 4.10 ($SD = 0.36$); the moderate group reported a mean of 4.83 ($SD = 0.14$), while the high motivation group had the highest mean at 5.41 ($SD = 0.27$). This pattern indicates that individuals with higher motivation tend to report greater motivation in their writing, particularly when compared to those in the low and moderate motivation groups.

An ordinal logistic regression was conducted to explore the relationship between SWAL and writing achievement motivation. The overall model was statistically significant ($\chi^2(100) = 18.23, p < .001$). The analysis revealed that SWAL was a strong predictor of writing achievement motivation ($B = 1.82, SE = 0.46, p < .001$). Specifically, each one-unit increase in SWAL resulted in a significantly higher likelihood of participants being placed in a higher motivation category.

Furthermore, the categorical comparison between motivation levels (low/moderate vs. moderate/high) had a significant effect ($B = 6.14, SE = 1.77, p = .001$). Participants in the low/moderate motivation category were substantially more likely to progress to a higher motivation group compared to those in the moderate/high category. On the other hand, individuals in the moderate/high group had a greater probability of advancing to a higher motivation level than those in the low/moderate group ($B = .66, SE = .14, p < .001$).

Table 3. Levels of Writing Achievement Motivation

| Groups of writing motivation | Mean | SD |
|---|---|---|
| Low | 4.10 | .36 |
| Moderate | 4.83 | .14 |
| High | 5.41 | .27 |

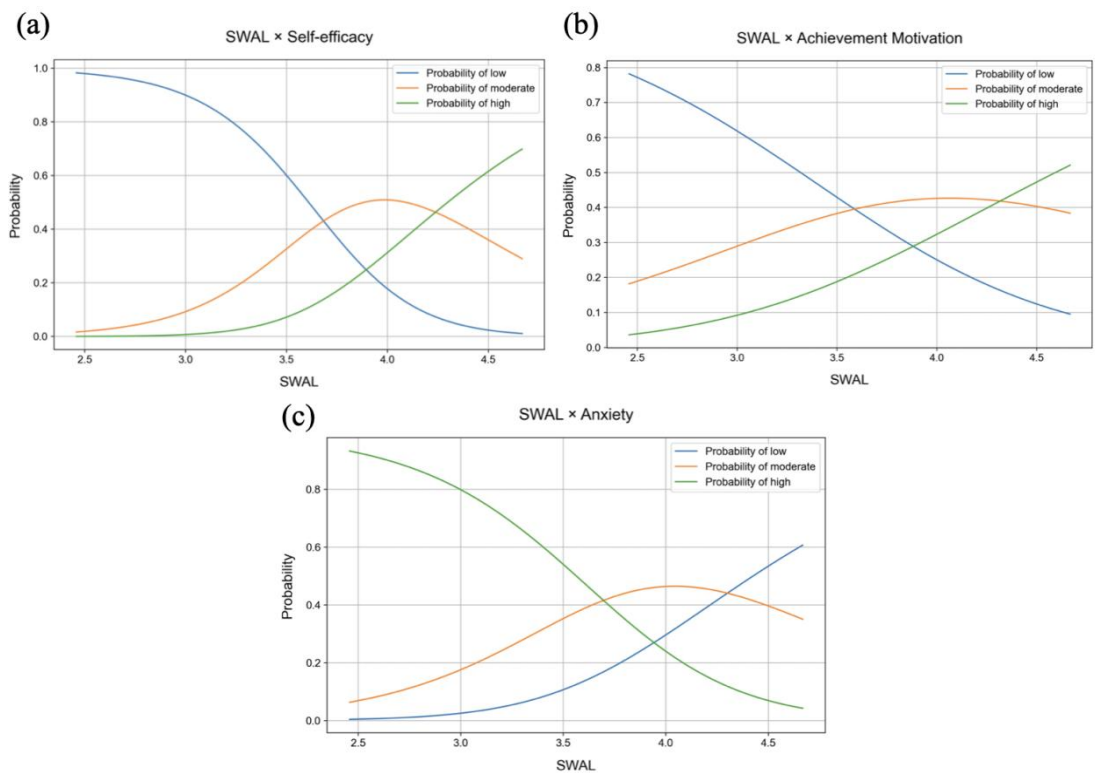

Figure 1. Prediction Probability Chart for SWAL predicting Writing Self-efficacy, Motivation, and Anxiety

Figure 1b. illustrated the probabilities of belonging to different motivation categories across varying SWAL levels. Participants with lower SWAL scores are primarily classified into the low motivation group. As SWAL scores increase, the likelihood of being in the moderate motivation category gradually rises, reaching its peak at mid-range SWAL levels. At the upper end of the SWAL spectrum, the probability of being categorized as having high motivation becomes dominant. These patterns visually emphasize the strong positive relationship between SWAL and motivation levels, as suggested by the regression analysis.

### 3.3 SWAL vs. writing anxiety

Table 4 provided an overview of the writing anxiety scores across groups with different levels. The mean motivation score for the low anxiety group was 4.10 ($SD = 0.36$), while the moderate group demonstrated a slightly higher mean score of 4.83

($SD$ = 0.14). The high anxiety group reported the highest mean motivation score of 5.41 ($SD$ = 0.27). These findings suggest a clear trend, where individuals with greater anxiety exhibit higher levels of anxiety compared to those in the low and moderate anxiety categories.

An ordinal logistic regression analysis was performed to investigate the relationship between SWAL and writing anxiety. The model demonstrated statistical significance ($\chi^2(1)$ = 45.13, $p$ < .001), indicating that SWAL is a significant predictor of writing anxiety. The results revealed that for each one-unit increase in SWAL, the likelihood of being classified into a higher level of writing anxiety decreased significantly ($B$ = -3.32, $SE$ = 0.59, $p$ < .001).

In addition, the comparisons between the anxiety levels (low/moderate and moderate/high) yielded notable findings. Specifically, participants in the low/middle anxiety category were much less likely to advance to a higher anxiety level compared to those in the moderate/high category ($B$ = -14.34, $SE$ = 2.44, $p$ < .001). Conversely, individuals transitioning from the middle to high anxiety groups showed a significant likelihood of progression, with a positive coefficient observed for this category ($B$ = 0.86, $SE$ = 0.14, $p$ < .001). These findings emphasize the strong association between SWAL and levels of writing anxiety, with SWAL playing a key role in reducing anxiety levels.

In Figure 1c, a visual representation of how SWAL levels influence the probabilities of falling into different writing anxiety categories was provided. At lower SWAL levels, individuals are most likely to belong to the high anxiety group, as indicated by the steep probability curve. As SWAL scores increase, the probability of being in the moderate anxiety group rises steadily, reaching its peak at mid-range SWAL levels. However, at higher SWAL levels, the probability of being classified in the low anxiety group surpasses the others, highlighting a negative association between SWAL and writing anxiety. This trend suggests that higher SWAL scores are linked to reduced writing anxiety, which is aligned with the regression analysis.

Table 4. Levels of Writing Anxiety

| Groups of writing anxiety | Mean | SD |
|---|---|---|
| Low | 1.57 | .21 |
| Moderate | 2.36 | .33 |
| High | 3.61 | .42 |

## 4. Discussion

This study employed ordinal logistic regression to examine the relationship between SWAL and three psychological factors: writing self-efficacy, writing achievement motivation, and writing anxiety. The results indicated that SWAL was positively associated with both writing self-efficacy and achievement motivation, while it showed a negative relationship with writing anxiety. Additionally, the predictive probability chart illustrated that higher SWAL scores were associated with lower levels of writing anxiety. The subsequent sections provide a detailed discussion of the potential factors underlying these findings.

### 4.1 Positive relationship between SWAL and writing self-efficacy

The statistical analysis revealed a strong positive relationship between SWAL and writing self-efficacy. This finding can be attributed to three key factors. First, students with higher writing assessment literacy are more familiar with assessment criteria, which can enhance their self-efficacy. In the SWAL scale, one dimension, labeled "knowledge," assesses students' understanding of assessment criteria or rubrics for L2 writing (Xu & Zheng, 2023; 2024). These rubrics typically outline expectations for English vocabulary, grammar, ideas, and other linguistic features. As a result, higher SWAL scores reflect a greater understanding of these criteria. Similarly, the self-efficacy scale includes two dimensions—ideas and conventions—that measure students' confidence in generating ideas and handling various linguistic aspects of writing, such as vocabulary and grammar (Bruning et al., 2013). Thus, higher SWAL scores are associated with higher self-efficacy scores.

Second, higher SWAL scores indicate that students possess more strategies and reflective experiences for assessing their own work, which can, in turn, enhance their self-efficacy. One dimension of the SWAL scale, labeled "behavior," serves two purposes. First, it examines whether students are able to apply strategies for different writing assessment tasks, such as using computer technology (Xu & Zheng, 2023; 2024). Second, it explores whether students engage in reflection on their writing assessment experiences to improve their writing proficiency. Higher SWAL scores thus suggest that students are more adept at using strategies and reflecting on their work. Research supports the idea that such strategies can boost self-efficacy. For example, Graham (2007) found that students who received strategy training and engaged in reflective writing exercises showed improvements in their self-efficacy. Similarly, studies have shown that learners with high self-efficacy in technology-based learning environments are more likely to put in effort when faced with learning challenges (Teng et al., 2021; Teng & Yang, 2023). These students are also more inclined to use metacognitive strategies to improve their writing skills. The availability of AI-assisted tools, which offer flexible online resources (Zhang, 2019) and provide feedback on language proficiency (Shen & Teng, 2024), further supports this process. Therefore, higher SWAL scores are positively correlated with higher self-efficacy scores.

Third, students with higher SWAL scores tend to have a more positive conception of feedback and greater engagement with both peer and teacher feedback, which in turn enhances their self-efficacy. One dimension of the SWAL scale, labeled "critique," serves two primary functions. First, it evaluates whether students can critically assess writing assessment results, such as feedback received. According to Brown et al. (2016), students' conceptions of feedback are crucial in determining whether learners believe feedback can help them achieve better outcomes. Their study found that the belief that "I use feedback" was positively associated with academic self-efficacy, suggesting that students who effectively use feedback are likely to experience higher self-efficacy.

The second purpose of the "critique" dimension is to assess students' engagement with peer and teacher feedback. Higher SWAL scores, therefore, indicate greater involvement in seeking and incorporating feedback from both peers and teachers. Research supports the idea that this engagement fosters self-efficacy. For instance, Bürgermeister et al. (2021) found that student teachers who repeatedly gave and received peer feedback reported higher self-efficacy. Similarly, Ruegg (2018) observed that students receiving teacher feedback showed a significant increase in writing self-efficacy compared to those who received only peer feedback. Moreover, Chan & Lam (2010) found that students who received summative feedback (which directs students towards achieving specific goals) experienced a larger decrease in self-efficacy compared to those who received formative feedback (which focuses on improvement and empowers students to achieve their goals). Taken together, these findings suggest that higher SWAL scores are positively correlated with higher self-efficacy scores, as students with higher SWAL scores are better able to critically engage with and use feedback to improve their writing.

### 4.2 *Positive relationship between SWAL and writing achievement motivation*

The current results revealed a strong positive relationship between SWAL and writing achievement motivation, which can be attributed to several factors. First, students with higher SWAL scores are more likely to believe that L2 writing assessment motivates students to write. This is evidenced by the "belief" dimension of the SWAL scale (Xu & Zheng, 2023, 2024), which explores how writing assessment impacts willingness to engage in such tasks. For instance, an item in the scale states, "L2 writing assessment can motivate students to write," reflecting the motivational role of assessment. Consequently, students with higher SWAL scores tend to exhibit stronger writing achievement motivation.

Another factor is the positive correlation between writing self-efficacy and writing achievement motivation. Students with higher SWAL scores often demonstrate higher self-efficacy, which has been shown to enhance motivation in writing tasks. According to Pajares (2003), learners with high writing self-efficacy scores exhibit greater writing achievement motivation. This relationship is further supported by research findings (Chea & Shumow, 2017; Sabti et al., 2019; Zhang & Guo, 2013), which consistently show significant positive correlations between writing self-efficacy and motivation. Thus, the strong link between SWAL and self-efficacy explains why students with higher SWAL scores also display higher motivation.

Finally, students with higher SWAL scores tend to adopt more effective strategies for mastering writing tasks, further enhancing their writing achievement motivation. The "behavior" dimension of the SWAL scale evaluates students' ability to apply strategies, such as utilizing computer technology for writing assessments. Research by Nasihah and Cahyono (2017) indicated that the use of appropriate learning strategies significantly improves writing achievement and motivation, as these strategies help students focus on task mastery. Additionally, the writing achievement motivation scale includes a dimension called "mastery goal," which emphasizes the development of competence and task mastery in English writing. The adoption of more learning strategies contributes to improved task mastery in key writing areas, such as vocabulary, organization, grammatical accuracy, and mechanics (Yusuf et al., 2019). Therefore, students with higher SWAL scores are better equipped with the strategies necessary for mastering writing tasks, which in turn boosts their motivation to succeed in writing.

*4.3 Negative relationship between SWAL and writing anxiety*

The results revealed a strong negative relationship between SWAL and writing anxiety, which can be explained by several key factors. Firstly, students with higher SWAL scores tend to experience more positive emotions, which can reduce their writing anxiety. The "belief" dimension of SWAL, which examines how writing assessments influence emotions, includes items such as "I am happy when the teacher assesses my writing" (Xu & Zheng, 2023, 2024). This suggests that higher SWAL scores are associated with more positive emotions (e.g., happiness), which in turn help alleviate somatic anxiety—defined by physiological symptoms like trembling or sweating (Young et al., 2019).

Secondly, students with higher SWAL scores are more likely to employ effective strategies for writing tasks, which also reduces anxiety. The "behavior" dimension of SWAL evaluates whether students use strategies such as computer technology to support their writing (Xu & Zheng, 2023, 2024). Research has shown that metacognitive writing strategies, which involve techniques for planning, monitoring, and evaluating writing, are negatively correlated with writing anxiety (Blasco, 2016). Additionally, AI-based strategies, such as using ChatGPT, have been found to reduce students' anxiety in English writing (Hawanti & Zubaydulloevna, 2023). Therefore, higher SWAL scores are linked to lower levels of writing anxiety.

Furthermore, students with higher SWAL scores tend to engage more with feedback from peers and teachers, which can further reduce writing anxiety. The "critique" dimension of SWAL assesses how students seek and incorporate feedback from both peers and teachers. Numerous studies (e.g., Bolourchi & Soleimani, 2021; Kurt & Atay, 2007; Yastıbaş & Yastıbaş, 2015) have shown that peer feedback significantly decreases writing anxiety. Similarly, teacher feedback, especially through digital platforms like blogs, is perceived as more useful and beneficial by students, which helps reduce anxiety (Abdullah et al., 2018).

Lastly, the negative correlation between writing self-efficacy, achievement motivation, and writing anxiety provides another explanation for the relationship between SWAL and writing anxiety. Studies have demonstrated that higher writing self-efficacy is associated with lower writing anxiety in both first and second languages (Sabti et al., 2019; Singh & Rajalingam, 2012). Additionally, anxiety is negatively correlated with achievement motivation (Csizér & Piniel, 2013; Sabti et al., 2019). This suggests that the strong link between SWAL and both self-efficacy and achievement motivation helps explain why students with higher SWAL scores experience lower levels of writing anxiety.

## 5. Limitations and implications in the future

While this study offers valuable insights into the relationship between SWAL and three key psychological factors in writing, there are several limitations that should be acknowledged. First, the study did not differentiate between high and low proficiency students, which could have provided a more nuanced understanding of how SWAL interacts with three key factors across different skill levels. Future research should explore whether these relationships vary based on students' writing proficiency, potentially highlighting different intervention strategies for diverse learner groups.

Second, the study employed ordinal logistic regression, which, while effective for examining the relationship between ordinal variables, does not allow for capturing potential non-linear interactions between the predictors and the outcome. Future studies may benefit from exploring more advanced techniques, such as structural equation modeling (SEM) or machine learning approaches, to better capture the complexities of these relationships.

Third, the study relied solely on data collected through a questionnaire, which may limit the depth and richness of the findings. While questionnaires are useful for collecting large-scale data, they do not capture the full range of student experiences, attitudes, or behaviors. Incorporating qualitative methods in the future, such as interviews or focus groups, would offer a more comprehensive understanding of how SWAL influences students' psychological factors related to writing.

The findings of this study have several important implications for second language (L2) writing instruction. First, the positive association between SWAL and writing self-efficacy and achievement motivation suggests that enhancing SWAL could be an effective intervention for fostering students' motivation and confidence in writing. Therefore, L2 teachers should aim to improve students' understanding of writing assessments by helping them grasp the purpose of writing evaluations, the potential emotional impacts, and the various methods used to assess writing. This can be done by providing clear explanations of assessment criteria and engaging students in discussions about how these assessments contribute to their writing development, which will foster a more positive view of writing assessments and encourage active participation.

Second, the negative relationship between SWAL and writing anxiety suggests that improving students' writing assessment literacy can help reduce anxiety in writing tasks. L2 teachers should focus on helping students understand the purpose of writing assessments and the different approaches to evaluating writing. By fostering a positive belief in writing assessments as tools for growth, teachers can motivate students to engage with assessments more confidently. Encouraging students to reflect on and use feedback constructively will also help them manage anxiety by giving them more control over their progress. Ultimately, enhancing students' SWAL can promote a more positive and less stressful writing experience.

# 6. Conclusion

This study examined the impact of SWAL on writing self-efficacy, achievement motivation, and writing anxiety. The results indicated that higher SWAL was positively associated with both writing self-efficacy and achievement motivation. Additionally, SWAL was negatively related to writing anxiety, indicating that greater assessment literacy can help reduce writing-related stress. These results have significant implications for second language (L2) writing instruction. By integrating SWAL training into the curriculum, L2 teachers can help students develop a better understanding of writing assessments, fostering greater motivation, confidence, and less anxiety in their writing tasks.